# HVS-BASED PERCEPTUAL COLOR COMPRESSION OF IMAGE DATA

*Lee Prangnell and Victor Sanchez*
{lee.prangnell, v.f.sanchez-silva}@warwick.ac.uk

Department of Computer Science, University of Warwick, England, UK

## ABSTRACT

In perceptual image coding applications, the main objective is to decrease, as much as possible, Bits Per Pixel (BPP) while avoiding noticeable distortions in the reconstructed image. In this paper, we propose a novel perceptual image coding technique, named Perceptual Color Compression (PCC). PCC is based on a novel model related to Human Visual System (HVS) spectral sensitivity and CIELAB Just Noticeable Color Difference (JNCD). We utilize this modeling to capitalize on the inability of the HVS to perceptually differentiate photons in very similar wavelength bands (e.g., distinguishing very similar shades of a particular color or different colors that look similar). The proposed PCC technique can be used with RGB (4:4:4) image data of various bit depths and spatial resolutions. In the evaluations, we compare the proposed PCC technique with a set of reference methods including Versatile Video Coding (VVC) and High Efficiency Video Coding (HEVC) in addition to two other recently proposed algorithms. Our PCC method attains considerable BPP reductions compared with all four reference techniques including, on average, 52.6% BPP reductions compared with VVC (VVC in All Intra still image coding mode). Regarding image perceptual reconstruction quality, PCC achieves a score of SSIM $\geq$ 0.99 in all tests in addition to a score of MS-SSIM $\geq$ 0.99 in all but one test. Moreover, MOS = 5 is attained in 75% of subjective evaluation assessments conducted.

*Index Terms* — Perceptual Image Coding, Visually Lossless Compression, Human Visual System, HEVC, VVC

## 1. INTRODUCTION

The primary objective in perceptual image coding research is to achieve visually lossless quality while simultaneously attaining the lowest possible bitstream size [1]. Regarding the modeling of perceptual image compression techniques, the Weber-Fechner law [2] confirms that there is a mathematical relationship between the subjective sensation of a physical stimulus and the intensity of the actual physical stimulus. This implies that there is a mathematical relationship between the perception of brightness and the intensity of physical luminance in nature. Moreover, it also implies that there is a mathematical relationship between perceived color (i.e., chroma, hue, saturation and contrast) and photon waves in nature. The scientific basis of the Weber-Fechner law is thus extremely useful for perceptual image coding research. Studies show that the HVS is relatively poor at detecting small differences in shades of a color [3, 4]. This could be in the form of slightly different shades of the same color or different colors that look very similar (e.g., aqua versus turquoise). In addition, the HVS is much more sensitive to the brightness of photons that are perceived as green; i.e., in the photon wavelength range of 500-585 nm [5]-[8] (see Fig. 1). This constitutes a cross-channel color masking effect in RGB data (i.e., the G channel is perceptually more important to the HVS). Therefore, the B and R channels are good targets for more aggressive compression levels; e.g., higher perceptual quantization levels applied to B and R data.

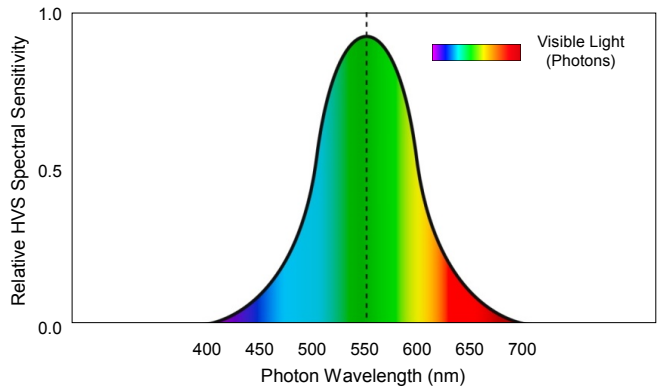

**Fig. 1**: The relative spectral sensitivity of the HVS to photons of various wavelengths (data: National Physical Laboratory [8]). The HVS is significantly more sensitive to the photon energies that the HVS interprets as the color green.

HEVC (ITU-T H.265) [9, 10] and VVC (ITU-T H.266) [11] are currently state-of-the-art video coding standards; these codecs possess still image coding capabilities. For example, HEVC and VVC include intra prediction coding tools [12, 13] and specialized profiles for the coding of RGB still image data [14, 15]. In terms of quantization, HEVC includes a scalar quantization method [16, 17] and VVC includes a vector (trellis-coded) quantization method [18]; as it stands, the main quantization methods in HEVC and VVC are not perceptually optimized. Other state-of-the-art techniques that are capable of direct RGB coding include FDPQ [19] and SPAQ [20], both of which are integrated into HEVC. FDPQ is a perceptual quantization method based on a Modulation Transfer Function (MTF) model; it operates at the transform coefficient level in R, G and B Transform Blocks (TBs). SPAQ is a perceptual quantization method designed to compress G data less coarsely according to how the HVS interprets photons perceived as green. Neither SPAQ nor FDPQ take JNCD into account.

In this paper, we propose a novel technique named Perceptual Color Compression (PCC). We employ a novel combined HVS spectral sensitivity and JNCD model that takes into account the way in which the HVS subjectively perceives photons in different wavelength bands. We implement PCC into HEVC as a Coding Block (CB)-level perceptual quantization algorithm; it is designed to separately adjust levels of quantization in R, G and B CBs. Furthermore, we integrate the CIELAB JNCD formula [21, 22] at the Coding Unit (CU) level to establish the JNCD threshold during the CB-level quantization process. In a nutshell, PCC perceptually quantizes data in the B and R CBs more coarsely than data in the G CBs. The aim is to avoid noticeable coding artifacts in the reconstructed CBs as well as achieving noteworthy BPP savings.

The rest of this paper is organized as follows. Section 2 includes an overview of lossy and perceptual image coding. We provide detailed information on the proposed PCC technique in Section 3. The evaluation, results and discussion of PCC are shown in Section 4. Finally, section 5 concludes the paper.

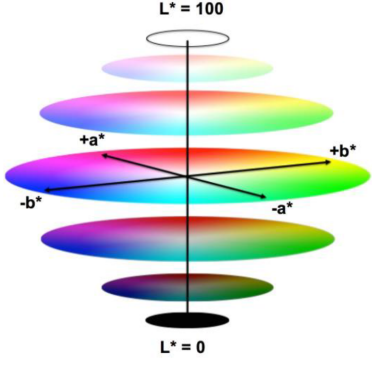

**Fig. 2**: Conceptual diagram of the CIELAB color space [21]. Note how CIELAB uses a 3D coordinate system, which separates the lightness of color from the chroma, hue and saturation of color.

## 2. PERCEPTUAL CODING: RELATED BACKGROUND

### 2.1 Scientific Background of Color-Based Perceptual Coding

In the context of JNCD-based modeling, studies confirm that the HVS is poor at distinguishing small differences in color [3, 4]. This has given rise to perceptually uniform color spaces including CIELAB [21, 22] (see Fig. 2). In terms of human perception and its relationship with perceptual image coding, an important concept to consider is the way in which the HVS interprets physical photons and physical luminance on VDUs and TVs (i.e., the perception of compressed signals displayed on VDUs and TVs). To this end, the HVS perceives the combination of photon waves and luminance as a vast range of colors. Note that color is purely a subjective phenomenon and thus does not exist outside of the perceptual domain. In terms of biology (photobiology), eye cone cell sensitivity experiments have revealed that 64% of such cells are sensitive to photons perceived as red, 32% perceived as green and 4% perceived as blue; this is known as trichromatic color vision [7]. Although there are more cones that are sensitive to photons interpreted as red, the HVS is much more sensitive to the perceived brightness of photons that are interpreted as green (see Fig. 1). For this reason, the green channel in the RGB color space is considered to be the most important channel in terms of color brightness perception [4] (see Fig. 1). To this end, we utilize a CIELAB JNCD threshold value to determine perceptually unnoticeable color differences in an image. Therefore, by employing a HVS spectral sensitivity (see Fig. 1) and CIELAB JNCD (see Fig. 2) model in PCC, we are able to account for the way in which the HVS interprets color difference and the brightness of color during the perceptual quantization process. For measuring CIELAB color difference, a variable is used known as Delta E, which is denoted as $\Delta E_{ab}$. The CIELAB $\Delta E_{ab}$ formula, as shown in [22, 23], is computed in (1):

$$\Delta E_{ab} = \sqrt{(L_2 - L_1)^2 + (a_2 - a_1)^2 + (b_2 - b_1)^2} \quad (1)$$

where $L$, $a$ and $b$ refer to Cartesian coordinates. In CIELAB, $L$ constitutes the lightness of a color; $L \in [0,100]$ (value 0 represents black and value 100 represents white). Coordinates $a$ and $b$ refer to the hue and saturation of a color. The chromatic $a$ axis extends from green to red; this is typically denoted as $(-a)$ for green and $(+a)$ for red. Likewise, the chromatic $b$ axis extends from blue to yellow, which is denoted as $(-b)$ for blue and $(+b)$ for yellow (see Fig. 2). The specific computations for $L$, $a$ and $b$ are shown in [24]. Note that CIELAB is a perceptually uniform color space that is RGB color space independent [22, 23]; therefore, CIELAB and $\Delta E_{ab}$ are very useful for perceptual image coding. Experiments in the field of colorimetry show that $\Delta E_{ab} \approx 2.3$ equates to the JNCD threshold [24].

---

**Algorithm 1**: Procedure for CB-Level QP Increments and Decrements

```
 1:  procedure Perceptual_CB_QP(Q_G, Q_B, Q_R)
 2:      ε << 1
 3:      if ΔE_ab < 2.3
 4:          while ΔE_ab < 2.3 do
 5:              CB-Level_QP_Incrementation:
 6:              repeat
 7:                  Increment_Blue_CB-Level_QP:  // Increment B CB QP first.
 8:                      i_B = 1; Q_B = (Q_B + i_B)
 9:                      do i_B++ until i_B = 6
10:                      if ΔE_ab = 2.3 ± ε, then goto End:
11:                      else goto Increment_Red_CB-Level_QP:
12:                  Increment_Red_CB-Level_QP:  // Increment R CB QP second.
13:                      i_R = 1; Q_R = (Q_R + i_R)
14:                      do i_R++ until i_R = 6
15:                      if ΔE_ab ± ε, then goto End:
16:                      else goto Increment_Green_CB-Level_QP:
17:                  Increment_Green_CB-Level_QP:  // Increment G CB QP last.
18:                      i_G = 1; Q_G = (Q_G + i_G)
19:                      do i_G++ until i_G = 3
20:                      if ΔE_ab = 2.3 ± ε, then goto End:
21:                      else goto CB-Level_QP_Incrementation:
22:              until ΔE_ab ∈ [2.3 − ε, 2.3 + ε]
23:              End:
24:          end while
25:      else
26:          while ΔE_ab > 2.3 do
27:              CB-Level_QP_Decrementation:
28:              repeat
29:                  Decrement_Green_CB-Level_QP:  // Decrement G CB QP first.
30:                      i_G = 1; Q_G = (Q_G − i_G)
31:                      do i_G−− until i_G = −3
32:                      if ΔE_ab = 2.3 ± ε, then goto End:
33:                      else goto Decrement_Red_CB-Level_QP:
34:                  Decrement_Red_CB-Level_QP:  // Decrement R CB QP second.
35:                      i_R = 1; Q_R = (Q_R − i_R)
36:                      do i_R−− until i_R = −6
37:                      if ΔE_ab = 2.3 ± ε, then goto End:
38:                      else goto Decrement_Blue_CB-Level_QP:
39:                  Decrement_Blue_CB-Level_QP:  // Decrement B CB QP last.
40:                      i_B = 1; Q_B = (Q_B − i_B)
41:                      do i_B−− until i_B = −6
42:                      if ΔE_ab = 2.3 ± ε, then goto End:
43:                      else goto CB-Level_QP_Decrementation:
44:              until ΔE_ab ∈ [2.3 − ε, 2.3 + ε]
45:              End:
46:          end while
47:      end if
48:  end procedure
```

### 2.2 Related Techniques and State-of-the-Art

Focusing on lossy RGB coding and quantization, the default quantization technique in HEVC is scalar Uniform Reconstruction Quantization (URQ) [16]. URQ uniformly quantizes coefficients in R, G and B Transform Blocks (TBs) at equal levels according to a quantization step size. VVC has replaced URQ with a trellis-coded Quantizer (TCQ) [18]. TCQ is a vector quantization technique that establishes the optimal quantization level for R, G and B transform coefficients. TCQ attains superior reconstruction quality compared with URQ. Both URQ in HEVC and TCQ in VVC are typically combined with RDOQ [25] to improve coding efficiency. Recent state-of-the-art perceptual quantization-based methods, which are capable of direct RGB image coding, include FDPQ [19] and SPAQ [20]. FDPQ quantizes transform coefficients according to a Modulation Transfer Function (MTF) model. High frequency coefficients are quantized more aggressively by FDPQ via the utilization of a Euclidean distance parameter. SPAQ is based on a spectral sensitivity model that operates according to how the HVS perceives color; in SPAQ, G data is quantized less coarsely. PCC improves upon all of the aforementioned methods (i.e., HEVC, VVC, FDPQ and SPAQ) by accounting for both JNCD and HVS spectral sensitivity.

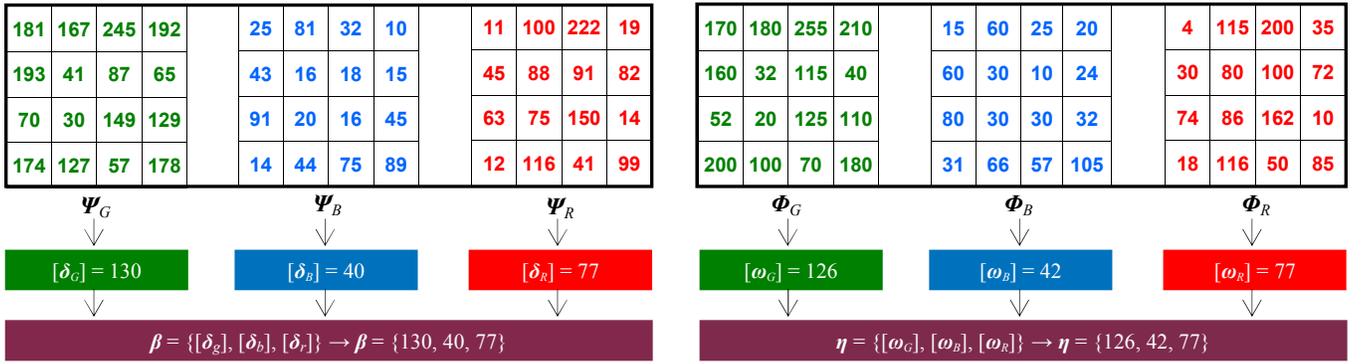

**Fig. 3**: CU-level JNCD computation. For brevity, we show dummy 4×4 raw G, B and R CBs ($\Psi_G$, $\Psi_B$ and $\Psi_R$) and dummy 4×4 reconstructed G, B and R CBs ($\Phi_G$, $\Phi_B$ and $\Phi_R$) within raw and reconstructed CUs, respectively. Though CUs and the constituent CBs in PCC are partitioned to the sample ranges of 8×8 to 64×64 [26], this figure is shown to illustrate the computation of the JNCD threshold (i.e., CIELAB $\Delta E_{ab} \approx 2.3$). This calculation is integral to PCC.

## 3. PROPOSED PCC TECHNIQUE

The direct coding of raw RGB data has been available to HEVC since the official standardization of the HEVC Range Extensions (RExt) [27]. PCC is thus integrated into JCT-VC HEVC HM RExt; our method exploits this direct RGB coding feature in HEVC HM RExt. PCC operates at the CB level for performing perceptual quantization operations and at the Coding Unit (CU) level for computing CIELAB color difference measurements. We employ the aforementioned CIELAB color difference formula, $\Delta E_{ab}$, as shown in (1) [24], to compute the JNCD threshold value. This is achieved by computing the color difference between raw pixels and reconstructed pixels at the CU level in the HEVC encoder loop. Recall that $\Delta E_{ab}$ constitutes the Euclidean distance between two colors in the CIELAB color space [21, 22]. A core objective in PCC is to adjust CB-level perceptual quantization until $\Delta E_{ab} \approx 2.3$ (i.e., the JNCD threshold).

As illustrated in Fig. 3, let $\Psi_G$, $\Psi_B$ and $\Psi_R$ denote the raw G, B and R CBs respectively, contained within a raw CU. Variables $\delta_G$, $\delta_B$ and $\delta_R$ refer to the mean G, B and R samples respectively, which are derived from $\Psi_G$, $\Psi_B$ and $\Psi_R$ respectively. Let $\Phi_G$, $\Phi_B$ and $\Phi_R$ denote the reconstructed G, B and R CBs respectively, contained within a reconstructed CU. Variables $\omega_G$, $\omega_B$ and $\omega_R$ correspond to the mean G, B and R CB samples respectively, which are derived from $\Phi_G$, $\Phi_B$ and $\Phi_R$ respectively. Next, let $\delta_G$, $\delta_B$ and $\delta_R$ be rounded and contained in set $\beta$. Also, let $\omega_G$, $\omega_B$ and $\omega_R$ be rounded and contained in set $\eta$. Sets $\beta$ and $\eta$ are thus treated as mean pixel values. To this end, $\beta = \{[\delta_G],[\delta_B],[\delta_R]\}$ and $\eta = \{[\omega_G],[\omega_B],[\omega_R]\}$. Note that $[\cdot]$ denotes the nearest integer function. The $\Delta E_{ab}$ formula in (1) is then employed to compare $\beta$ and $\eta$. The aim with comparing $\beta$ and $\eta$ is to establish if $\Delta E_{ab} \approx 2.3$, as detailed in Algorithm 1. As per the while loop in line 4 of Algorithm 1, if $\Delta E_{ab} < 2.3$ when comparing $\beta$ with $\eta$, PCC increments CB-level perceptual QPs until $\Delta E_{ab} \approx 2.3$. The aforementioned HVS-based spectral sensitivity modeling, which is based on the model illustrated in Fig. 1, dictates the order in which the CB-level perceptual QPs are incremented. The B CB QP is incremented first because B is the least perceptually important color channel. On the other hand, G is considered as the most perceptually important color channel. Therefore, the G CB QP is incremented last. This incrementation is repeated until $\Delta E_{ab} \approx 2.3$. Conversely, and as shown in the while loop in line 26 of Algorithm 1, if $\Delta E_{ab} > 2.3$ when comparing $\beta$ with $\eta$, this equates to the fact that the JNCD threshold has been exceeded. When this occurs, PCC decreases CB-level QPs. More specifically, the G CB QP is decremented first and the B CB QP is decremented last. This process is repeated until $\Delta E_{ab} \approx 2.3$. Fig. 3 shows an illustration of a dummy JNCD computation using $\beta$ and $\eta$. With the toy G, B and R sample values used in Fig. 3, the mean values in set $\beta$ equate to a purple-like color. Moreover, the mean values in set $\eta$ equate to an imperceptibly different shade of the color derived from $\beta$. In this example, $\Delta E_{ab} \approx 2.3$ (i.e., the JNCD threshold has been reached).

In the proposed PCC algorithm, the CB-level perceptual QPs, denoted as $Q_G$, $Q_B$ and $Q_R$, and the corresponding CB-level Quantization Step Sizes (QSteps), denoted as $S_G$, $S_B$ and $S_R$, are shown in (2)-(7), respectively:

$$Q_G(S_G) = \left(\left\lceil 6 \times \log_2(S_G) \right\rceil + 4 \pm i_G \right) \quad (2)$$

$$S_G(Q_G) = 2^{\frac{Q_G - 4 \pm i_G}{6}} \quad (3)$$

$$Q_B(S_B) = \left(\left\lceil 6 \times \log_2(S_B) \right\rceil + 4 \pm i_B \right) \quad (4)$$

$$S_B(Q_B) = 2^{\frac{Q_B - 4 \pm i_B}{6}} \quad (5)$$

$$Q_R(S_R) = \left(\left\lceil 6 \times \log_2(S_R) \right\rceil + 4 \pm i_R \right) \quad (6)$$

$$S_R(Q_R) = 2^{\frac{Q_R - 4 \pm i_R}{6}} \quad (7)$$

where $i_G$, $i_B$ and $i_R$ refer to incremental or decremental values for increasing or decreasing, respectively, the G CB, B CB and R CB QPs until $\Delta E_{ab} \approx 2.3$. The $\Delta E_{ab} \approx 2.3$ approximation is checked by ensuring that $\Delta E_{ab} \in [2.3 - \varepsilon, 2.3 + \varepsilon]$, where $\varepsilon \ll 1$. This check takes place on line 22 and on line 44 in Algorithm 1. Note that if $\Delta E_{ab} < 2.3$, then $i_G = 1$, $i_B = 1$ and $i_R = 1$. Otherwise, if $\Delta E_{ab} > 2.3$, then $i_G = -1$, $i_B = -1$ and $i_R = -1$. PCC signals the CB-level QP offset data to the decoder via the Picture Parameter Set (PPS), which is available in HEVC HM RExt [27]. Note that the signaling of CB-level QP offsets in the PPS allows for a straightforward encoder side implementation of PCC.

## 4. EVALUATION, RESULTS AND DISCUSSION

The proposed PCC method is implemented into HEVC HM RExt 16.7 [28]. In the experiments, PCC is configured to use All Intra coding and the main_444_intra profile for RGB still image coding. In the evaluations, we compare PCC with two recently proposed perceptual compression methods: namely, FDPQ [19] and SPAQ [20]. We also compare PCC with the JVET Versatile Video Coding (VVC) VTM 10.0 codec [29] and the HEVC HM RExt 16.20 codec [30]. VVC VTM 10.0 and HEVC HM RExt 16.20 are configured to use All Intra coding and the main_444_intra profile. PCC and the four reference techniques are tested on 12 raw RGB 4:4:4 images, all of which are HD 1080p in resolution size except for the Computerized Tomography (CT) image (1000×1000) and the Whole Slide Image (WSI), which is 4K in resolution size.

**Table 1**: Tabulated Bits Per Pixel (BPP), SSIM, MS-SSIM and MOS results for the proposed PCC method and reference techniques SPAQ, FDPQ, VVC and HEVC. For the proposed PCC technique, note that negative results — compared with the reference techniques — are shown in red text.

| | Bits Per Pixel (BPP), SSIM and MS-SSIM Scores and MOS for Proposed PCC Method versus Reference Techniques | | | | | | | | | | | | | | | | | | | |
|---|---|---|---|---|---|---|---|---|---|---|---|---|---|---|---|---|---|---|---|---|
| | **Bits Per Pixel (BPP)** | | | | | **SSIM** | | | | | **MS-SSIM** | | | | | **MOS (Rounded)** | | | | |
| RGB Data | PCC | SPAQ | FDPQ | VVC | HEVC | PCC | SPAQ | FDPQ | VVC | HEVC | PCC | SPAQ | FDPQ | VVC | HEVC | PCC | SPAQ | FDPQ | VVC | HEVC |
| **BirdsInCage** | 0.40 | 0.60 | 1.00 | 1.05 | 1.10 | 0.99 | 0.99 | 0.99 | 0.99 | 0.99 | 0.99 | 0.99 | 0.99 | 0.99 | 0.99 | 5 | 5 | 5 | 5 | 5 |
| **Bubbles** | 0.51 | 0.82 | 1.17 | 1.04 | 1.11 | 0.99 | 0.99 | 0.99 | 0.99 | 0.99 | 0.99 | 0.99 | 0.99 | 0.99 | 0.99 | 5 | 5 | 5 | 5 | 5 |
| **CrowdRun** | 2.16 | 3.92 | 5.03 | 5.57 | 5.85 | 0.99 | 0.99 | 0.99 | 0.99 | 0.99 | 0.99 | 0.99 | 0.99 | 0.99 | 0.99 | 5 | 5 | 5 | 5 | 5 |
| **CT** | 0.32 | 0.36 | 0.59 | 0.44 | 0.47 | 0.99 | 0.99 | 0.99 | 0.99 | 0.99 | 0.99 | 0.99 | 0.99 | 0.99 | 0.99 | 5 | 5 | 5 | 5 | 5 |
| **DucksAndLegs** | 2.21 | 3.79 | 4.69 | 5.38 | 5.55 | 0.99 | 0.99 | 0.99 | 0.99 | 0.99 | 0.99 | 0.99 | 0.99 | 0.99 | 0.99 | 5 | 5 | 5 | 5 | 5 |
| **Kimono** | 0.50 | 1.01 | 1.87 | 1.82 | 1.90 | 0.99 | 0.99 | 0.99 | 0.99 | 0.99 | 0.99 | 0.99 | 0.99 | 0.99 | 0.99 | 5 | 5 | 5 | 5 | 4 |
| **OldTownCross** | 1.30 | 3.33 | 4.57 | 5.26 | 5.49 | 0.99 | 0.99 | 0.99 | 0.99 | 0.99 | 0.98 | 0.99 | 0.99 | 0.99 | 0.99 | 5 | 5 | 5 | 5 | 5 |
| **ParkScene** | 1.10 | 2.35 | 3.33 | 3.58 | 3.78 | 0.99 | 0.99 | 0.99 | 0.99 | 0.99 | 0.99 | 0.99 | 0.99 | 0.99 | 0.99 | 4 | 5 | 5 | 5 | 5 |
| **Seeking** | 1.71 | 3.74 | 4.81 | 5.53 | 5.80 | 0.99 | 0.99 | 0.99 | 0.99 | 0.99 | 0.99 | 0.99 | 0.99 | 0.99 | 0.99 | 4 | 5 | 5 | 5 | 5 |
| **Traffic** | 1.14 | 1.35 | 1.95 | 1.83 | 2.03 | 0.99 | 0.99 | 0.99 | 0.99 | 0.99 | 0.99 | 0.99 | 0.99 | 0.99 | 0.99 | 5 | 5 | 5 | 5 | 5 |
| **VenueVu** | 0.64 | 0.73 | 1.07 | 0.92 | 1.01 | 0.99 | 0.99 | 0.99 | 0.99 | 0.99 | 0.99 | 0.99 | 0.99 | 0.99 | 0.99 | 4 | 4 | 4 | 4 | 4 |
| **WSI (4K)** | 0.53 | 0.52 | 0.78 | 0.63 | 0.66 | 0.99 | 0.99 | 0.99 | 0.99 | 0.99 | 0.99 | 0.99 | 0.99 | 0.99 | 0.99 | 5 | 5 | 5 | 5 | 5 |

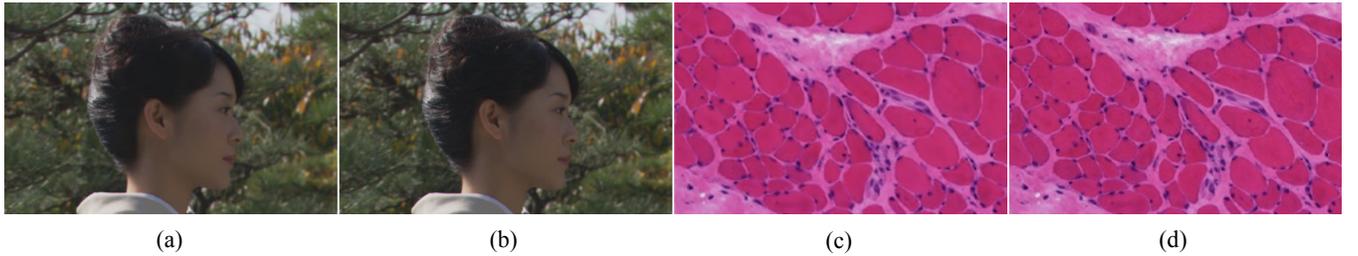

(a)      (b)      (c)      (d)

**Fig. 4**: (a) PCC-coded Kimono RGB image (HD 1080p) versus (b) Kimono raw RGB image. PCC attains an overall MOS = 5, SSIM ≥ 0.99 in addition to MS-SSIM ≥ 0.99 in the Kimono evaluations. (c) PCC-coded Whole Slide Image (WSI) 4K RGB image versus (d) WSI raw RGB image (4K). PCC also achieves an overall MOS = 5, SSIM ≥ 0.99 as well as MS-SSIM ≥ 0.99 in the WSI evaluations.

In terms of assessing the rate-distortion performance of PCC compared with the reference techniques, we utilize both SSIM [31] and MS-SSIM [32] to measure the perceptual reconstruction quality of all coded images. The primary objective with PCC is to achieve the lowest possible BPP without incurring perceptually conspicuous compression artifacts in the coded images (i.e., to achieve visually lossless quality). In the experimental setup, an initial frame-level QP (iQP) is applied to all reference techniques in order to target scores of SSIM ≥ 0.99 and also MS-SSIM ≥ 0.99. Note that scores of SSIM ≥ 0.99 and MS-SSIM ≥ 0.99 usually equates to a subjective evaluation Mean Opinion Score (MOS) = 5. MOS = 5 typically constitutes visually lossless quality [33, 34]. To ensure that fair testing is achieved in the evaluation, the same iQP that is used for the reference techniques is applied to PCC (in all tests); the objective here is to establish if PCC-coded images achieve scores of SSIM ≥ 0.99 and MS-SSIM ≥ 0.99. In the subjective evaluations, we follow the conditions specified in ITU-T Rec. P.910 [35]. ITU-T Rec. P.910 advises a minimum of four participants, a viewing distance of approximately 0.75m and the use of MOS to grade perceptual image quality (range: MOS = 1 to MOS = 5). We undertake the subjective evaluations to ascertain if PCC-coded images achieve visually lossless quality.

PCC attains considerable BPP reductions in the vast majority of tests (see Table 1). Taking into account the SSIM result, the MS-SSIM result, the MOS attained in the subjective evaluations and the BPP savings achieved, the best overall result was obtained on the Kimono sequence (see Fig. 4). In this particular test, 72.5% BPP savings (versus VVC) and 50.4% BPP savings (versus SPAQ) were accomplished by PCC. Regarding perceptual visual quality assessments, the proposed PCC technique achieves a score of SSIM ≥ 0.99 and a score of MS-SSIM ≥ 0.99 in all tests except for the OldTownCross test, in which MS-SSIM ≥ 0.98 was attained (see Table 1). Recall that perceptual reconstruction quality scores of SSIM ≥ 0.99 and MS-SSIM ≥ 0.99 typically indicate visually lossless quality for the compressed picture [33, 34].

Ten individuals participated in the subjective evaluations, in which a total of 600 visual comparisons were carried out. All techniques, including PCC, were compared with the raw data. On average, PCC achieves an MOS = 5 in 75% of tests (see Table 1). In other words, the participants were unable to notice any differences between PCC-coded images and the raw data in 75% of tests; see Fig. 4 for visual examples. Near-visually lossless quality (i.e., MOS = 4) was reported in 25% tests; that is, compression artifacts were very slightly perceptible in PCC-coded images. Among other things, the evaluation proves that there is a strong correlation between both the SSIM and MS-SSIM perceptual metrics and the subjective evaluation MOS. We discovered that images with lower color variance (i.e., CT, Bubbles and WSI) attained a higher amount of MOS = 5 results overall (see Table 1) for both PCC and the reference techniques. In terms of BPP savings (% gains), PCC performed much better than the reference techniques on images with a higher color variance (e.g., Kimono and OldTownCross). In terms of encoding and decoding time performances, there are no differences between PCC, SPAQ, FDPQ and HEVC. However, PCC is considerably faster than VVC.

## 5. CONCLUSION

We have proposed a color-based perceptual image coding technique, named PCC, for application with RGB image data. In PCC, we exploit HVS spectral sensitivity and JNCD-based modeling in order to guide perceptual quantization adjustments at the CB level. Compared with the reference techniques, PCC attains vast BPP savings in all tests; the largest BPP saving (76.4% reductions), was recorded in the OldTownCross test. Furthermore, PCC achieves visually lossless quality, which was confirmed by the subjective evaluations (participants recorded an MOS = 5 in 75% of tests). In addition, PCC attains perceptual reconstruction quality scores of SSIM ≥ 0.99 (in all tests) and MS-SSIM ≥ 0.99 (in all but one test).

## REFERENCES

[1] H. R. Wu, A. R. Reibman, W. Lin, F. Pereira, and S. S. Hemami, "Perceptual Visual Signal Compression and Transmission," *Proc. IEEE*, vol. 101, no. 9, pp. 2025-2043, 2013.

[2] G. T. Fechner, "Elements of Psychophysics, Volume 1," (Translated by H.E. Adler), New York: Holt, Rinehart & Winston, 1860.

[3] A. Chaparro, C. F. Stromeyer, E. P. Huang, R. E. Kronauer and R. T. Eskew, Jr., "Colour is what the eye sees best," *Nature*, vol. 361, pp. 348-350, 1993.

[4] K. R. Gegenfurtner, "Cortical Mechanisms of Colour Vision," *Nature Neuroscience*, vol. 4, pp. 563-572, 2003.

[5] H-J. Lewerenz, "Photons in Natural and Life Sciences," Berlin, Germany: Springer, 2012.

[6] G. Glaeser and H. F. Paulus, "The Evolution of the Eye," Vienna, Austria: Springer, 2014.

[7] L. O. Björn, "Photobiology," Guangzhou, China: Springer, 2015.

[8] National Physical Laboratory. Relative HVS Spectral Sensitivity Data. Available: http://www.npl.co.uk/

[9] ITU-T: Rec. H.265/HEVC (Version 5) | ISO/IEC 23008-2, Information Technology – Coding of Audio-visual Objects, *JCT-VC (ITU-T/ISO/IEC)*, 2018.

[10] G. Sullivan, J-R. Ohm, W. Han and T. Wiegand, "Overview of the High Efficiency Video Coding (HEVC) Standard," *IEEE Trans. Circuits Syst. Video Technol.*, vol. 22, no. 12, pp. 1649-1668, 2012.

[11] ITU-T: Rec. H.266/VVC (Version 1) | ISO/IEC 23090-3, Information Technology – Coding of Audio-visual Objects, *JVET (ITU-T/ISO/IEC)*, 2020.

[12] J. Lainema, F. Bossen, W-J. Han, J. Min and K. Ugur, "Intra Coding of the HEVC Standard," *IEEE Trans. Circuits Syst. Video Technol.*, vol. 22, no. 12, pp. 1792-1801, 2012.

[13] B. Bross, J. Chen, S. Liu and Y. K. Wang, "Versatile Video Coding (Draft 7)," *document JVET-P2001*, 16th JVET meeting: Geneva, CH, 1-11 Oct. 2019.

[14] Joint Collaborative Team on Video Coding (JCT-VC). JCT-VC HEVC HM RExt Software Manual [Online]. Available: https://hevc.hhi.fraunhofer.de/

[15] Joint Video Experts Team (JVET) VVC VTM Software Manual [Online]. Available: https://jvet.hhi.fraunhofer.de/

[16] M. Budagavi, A. Fuldseth, G. Bjøntegaard, V. Sze and M. Sadafale, "Core Transform Design in the High Efficiency Video Coding (HEVC) Standard," *IEEE J. Sel. Topics Signal Process.*, vol. 7, no. 6, pp. 1649-1668, 2013.

[17] M. Wein, "Quantizer Design," in *High Efficiency Video Coding: Coding Tools and Specification*, Springer, 2015, pp. 213-214.

[18] H. Schwarz, T. Nguyen, D. Marpe and T. Wiegand, "Hybrid Video Coding with Trellis-Coded Quantization," *Data Compression Conf.*, Snowbird, Utah, USA, 2019, pp. 182-191.

[19] L. Prangnell and V. Sanchez, "Frequency-Dependent Perceptual Quantisation for Visually Lossless Compression Applications," *arXiv:1906.03395 [cs.MM]*, 2019.

[20] L. Prangnell and V. Sanchez, "Spatiotemporal Adaptive Quantization for the Perceptual Video Coding of RGB 4:4:4 Data," *arXiv:2005.07928 [cs.MM]*, 2020.

[21] Sappi Limited, "Defining and Communicating Color: The CIELAB System," Technical Document, pp. 5, 2013.

[22] A. R. Robertson, "Historical development of CIE recommended color difference equations," *Color Research and Application*, vol. 15, no. 3, pp. 167-170, 1990.

[23] M. R. Luo, G. Cui and B. Rigg, "The development of the CIE 2000 colour-difference formula: CIEDE2000," *Color Research and Application*, vol. 26, no. 5, pp. 340-350, 2001.

[24] G. Sharma, R. Bala, "Color Fundamentals for Digital Imaging," in *Digital Color Imaging Handbook*, CRC Press, 2002, pp. 31.

[25] M. Karczewicz, Y. Ye and I. Chong, "Rate Distortion Optimized Quantization," in *VCEG-AH21 (ITU-T SG16/Q6 VCEG)*, Antalya, Turkey, 2008.

[26] I-K. Kim, J. Min, T. Lee, W-J. Han and J. Park, "Block Partitioning Structure in the HEVC Standard," *IEEE Trans. Circuits Syst. Video Technol.*, vol. 22, no. 12, pp. 1697-1706, 2012.

[27] D. Flynn, D. Marpe, M. Naccari, T. Nguyen, C. Rosewarne, K. Sharman, J. Sole and J. Xu, "Overview of the Range Extensions for the HEVC Standard: Tools, Profiles, and Performance," *IEEE Trans. Circuits Syst. Video Techn., vol.* 26, no. 1, pp. 4-19, 2016.

[28] Joint Collaborative Team on Video Coding (JCT-VC). JCT-VC HEVC HM Reference Software, HEVC HM 16.7 [Online]. Available: hevc.hhi.fraunhofer.de/

[29] Joint Video Experts Team (JVET). JVET VVC VTM Reference Software, VVC VTM 10.0 [Online]. Available: https://jvet.hhi.fraunhofer.de/

[30] Joint Collaborative Team on Video Coding (JCT-VC). JCT-VC HEVC HM Reference Software, HEVC HM 16.20 [Online]. Available: hevc.hhi.fraunhofer.de/

[31] Z. Wang, A. C. Bovik, H. R. Sheikh, and E. P. Simoncelli, "Image Quality Assessment: From Error Visibility to Structural Similarity," *IEEE Trans. Image Processing*, vol. 13, no. 4, pp. 600-612, 2004.

[32] Z. Wang, E. P. Simoncelli, A. C. Bovik, "Multiscale Structural Similarity for Image Quality Assessment," *IEEE Asilomar Conf. Signals, Systems and Computers*, Pacific Grove, CA, USA, 2003, pp. 1398-1402.

[33] T. Zinner, O. Hohlfeld, O. Abboud, T. Hossfeld, "Impact of frame rate and resolution on objective QoE metrics," *IEEE Int. Works. Quality of Multimedia Experience*, Trondheim, Norway, 2010, pp. 29-34.

[34] M. Zanforlin, D. Munaretto, A. Zanella and M. Zorzi, "SSIM-based video admission control and resource allocation algorithms," *Int. Symp. Modeling and Optimization in Mobile, Ad Hoc, and Wireless Networks*, Hammamet, Tunisia, 2014, pp. 656-661.

[35] ITU-T, "Rec. P.910: Subjective video quality assessment methods for multimedia applications," *ITU-T*, 2008.